\documentclass[preprint]{aastex}

\newcommand{\yr}{{\rm\,yr}}

\newcommand{\au}{{\rm\,AU}}

\begin{document}

\title{Planetary Migration and Eccentricity and Inclination Resonances
       in Extrasolar Planetary Systems}

\author{Man Hoi Lee\altaffilmark{1,2} and
        Edward W. Thommes\altaffilmark{3}}
\altaffiltext{1}{Department of Physics, University of California,
                 Santa Barbara, CA 93106.}
\altaffiltext{2}{Department of Earth Sciences and Department of
                 Physics, University of Hong Kong, Pokfulam Road,
                 Hong Kong.}
\altaffiltext{3}{Department of Physics, University of Guelph, Guelph,
                 ON N1G 2W1, Canada.}

\begin{abstract}
The differential migration of two planets due to planet-disk
interaction can result in capture into the 2:1 eccentricity-type
mean-motion resonances.
Both the sequence of 2:1 eccentricity resonances that the system is
driven through by continued migration and the possibility of a
subsequent capture into the 4:2 inclination resonances are sensitive
to the migration rate within the range expected for type II migration
due to planet-disk interaction.
If the migration rate is fast, the resonant pair can evolve into a
family of 2:1 eccentricity resonances different from those found by
\citet{lee04}.
This new family has outer orbital eccentricity $e_{2} \ga 0.4$--$0.5$,
asymmetric librations of both eccentricity resonance variables, and
orbits that intersect if they are exactly coplanar.
Although this family exists for an inner-to-outer planet mass ratio
$m_{1}/m_{2} \ga 0.2$, it is possible to evolve into this family by
fast migration only for $m_{1}/m_{2} \ga 2$.
\citet{tho03} have found that a capture into the 4:2 inclination
resonances is possible only for $m_{1}/m_{2} \la 2$.
We show that this capture is also possible for $m_{1}/m_{2} \ga 2$ if
the migration rate is slightly slower than that adopted by Thommes \&
Lissauer.
There is significant theoretical uncertainty in both the sign and the
magnitude of the net effect of planet-disk interaction on the orbital
eccentricity of a planet.
If the eccentricity is damped on a timescale comparable to or shorter
than the migration timescale, $e_{2}$ may not be able to reach the
values needed to enter either the new 2:1 eccentricity resonances or
the 4:2 inclination resonances.
Thus, if future observations of extrasolar planetary systems were to
reveal certain combinations of mass ratio and resonant configuration,
they would place a constraint on the strength of eccentricity damping
during migration, as well as on the rate of the migration itself.
\end{abstract}

\section{INTRODUCTION}
\label{intro}

Extrasolar planet searches have to date yielded about 33 systems with
multiple planets, and at least 8 of these systems have a pair of
planets known or suspected to be in mean-motion resonances.
It is well established that the outer two planets in the GJ 876 system
are deep in 2:1 resonances, with the retrograde periapse precessions
induced by the 2:1 resonances having been observed for more than one
full period \citep{mar01,lau01,riv01,lee02,lau05,riv05}.
In the GJ 876 system, both of the lowest order, eccentricity-type
mean-motion resonance variables
\begin{eqnarray}
\theta_1 &=& \lambda_1 - 2 \lambda_2 + \varpi_1 \label{theta1} \\
\theta_2 &=& \lambda_1 - 2 \lambda_2 + \varpi_2 , \label{theta2}
\end{eqnarray}
and hence the secular apsidal resonance variable
\begin{equation}
\theta_{\rm SAR} = \varpi_1 - \varpi_2 = \theta_1 - \theta_2 ,
\end{equation}
librate about $0^\circ$, which mean that the periapses are nearly
aligned and that conjunctions of the planets occur when both planets
are near periapse.
In the above equations, $\lambda_1$ and $\lambda_2$ are the mean
longitudes of the inner and outer planets, respectively, and
$\varpi_j$ are the longitudes of periapse.
There are three other systems with planets in 2:1 resonances: HD 82943
\citep{may04,fmb05,lee06,bgfm08}, HD 128311 \citep{vog05}, and
HD 73526 \citep{tin06,san07}, although it should be noted that a pair
of planets in 1:1 resonance is a plausible alternative for at least
HD 82943 and HD 128311 \citep{goz06}.
In addition, the HD 45364 \citep{cor09}, 55 Cancri
\citep{mar02,mca04}, HD 60532 \citep{des08,las09}, and HD 202206
\citep{cor05} systems have planets that are in 3:2, 3:1, 3:1 and 5:1
resonances, respectively.
There are uncertainties in converting this data into the fraction of
multiple-planet systems with mean-motion resonances.
Some of the suspected resonant pairs may not be confirmed eventually
(see, e.g., \citealt{fis08} for 55 Cancri).
On the other hand, the number of resonant pairs that remain undetected
could be quite large, because the radial velocity variation due to two
planets in resonance (in particular, 2:1) could be indistinguishable
from that due to a single planet for certain planetary mass ratio and
orbital eccentricities, given the precision levels of the existing
radial velocity surveys \citep{ang08,giu09}.
Nevertheless, the existing data indicate that $\sim 20\%$ of
multiple-planet systems have mean-motion resonances.

Mean-motion resonances can be easily established during planet
formation by the convergent migration of planets due to interactions
with the circumstellar gas disk.
Two giant planets that are massive enough to open gaps in the disk
individually can clear out the disk material between them rather
quickly, and the outer planet is forced to migrate inward by the disk
material outside its orbit (and the inner planet outward if there is
any disk material left inside its orbit) \citep{bry00,kle00}.
Both hydrodynamic and three-body simulations (with imposed migration
for the latter) have shown that the convergence of the orbits
naturally leads to capture into mean-motion resonances
\citep{bry00,kle00,sne01,lee02,nel02,pap03,tho03,kle04,kle05,lee04}.

The ubiquity of mean-motion resonances in extrasolar planetary systems
and the ease of capture into such resonances by convergent migration
have prompted investigations into the variety of stable mean-motion
resonance configurations
\citep{lee02,lee03,bfm03,bmf06,fbm03,had03,ji03,tho03,lee04,voy05,
voy06,mar06,mbf06}.
The 2:1 mean-motion commensurability has received the most attention,
because it is the most common one observed and includes the best case,
GJ 876.
With the exception of \citet{tho03}, all of the works just cited have
focused on systems with two planets on {\it coplanar} orbits.
For small orbital eccentricities, antisymmetric configurations with
$\theta_1$ librating about $0^\circ$ and $\theta_2$ about $180^\circ$
(as in the case of the Jovian satellites Io and Europa) are the only
stable 2:1 resonance configuration with both $\theta_1$ and $\theta_2$
librating.
For moderate to large eccentricities, the Io-Europa configuration is
not stable, but there is a wide variety of other stable 2:1 resonance
configurations, including symmetric configurations with both
$\theta_1$ and $\theta_2$ librating about $0^\circ$ (as in the GJ 876
system), asymmetric configurations with $\theta_1$ and $\theta_2$
librating about angles other than $0^\circ$ and $180^\circ$ (some with
intersecting orbits), and antisymmetric configurations with $\theta_1
\approx 180^\circ$ and $\theta_2 \approx 0^\circ$ (and intersecting
orbits).\footnote{
Throughout this paper, we often use ``$\theta_1 \approx x$'' as an
abbreviation for ``the libration of $\theta_1$ about an angle $x$''
(and similarly for the other resonance variables) when we describe a
resonance configuration.
}
\citet{lee04} has shown that the sequence of 2:1 resonance
configurations that a system with initially coplanar and nearly
circular orbits is driven through by continued migration depends
mainly on the planetary mass ratio $m_1/m_2$, if the migration rate is
sufficiently slow.
However, there are stable 2:1 resonance configurations (e.g., those
with $\theta_1 \approx 180^\circ$ and $\theta_2 \approx 0^\circ$) that
cannot be reached by the convergent migration of planets with constant
masses and initially coplanar and nearly circular orbits.
If real systems with these configurations are ever found, their origin
would require a change in the planetary mass ratio $m_1/m_2$ during
migration, multiple-planet scattering in crowded planetary systems, or
a migration scenario involving inclination resonances \citep{lee04}.

\citet{tho03} have studied the convergent migration of planets with
{\it non-coplanar} orbits and found that, subsequent to the capture
into the 2:1 eccentricity resonances, a capture into the 4:2
inclination resonances (which are the lowest order inclination
resonances at the 2:1 commensurability) is possible if $m_1/m_2 \la
2$.
The 4:2 inclination-type mean-motion resonance variables are
\begin{eqnarray}
\phi_{11} &=& 2 \lambda_1 - 4 \lambda_2 + 2 \Omega_1 \label{phi11} \\
\phi_{22} &=& 2 \lambda_1 - 4 \lambda_2 + 2 \Omega_2 , \label{phi22}
\end{eqnarray}
where $\Omega_j$ are the longitudes of the ascending node.
The simultaneous librations of $\phi_{11}$ and $\phi_{22}$ mean that
the mixed resonance variable
\begin{equation}
\phi_{12} = 2 \lambda_1 - 4 \lambda_2 + \Omega_1 + \Omega_2
          = (\phi_{11} + \phi_{22})/2
\end{equation}
also librates.
As the system enters the inclination resonances, the mutual
inclination of the orbits can grow rapidly to tens of degrees.
In some cases, the system eventually evolves out of the inclination
resonances, and the eccentricity resonances switch to the $\theta_1
\approx 180^\circ$ and $\theta_2 \approx 0^\circ$ configurations
mentioned above.
\citet{tho03} showed an example with $m_1/m_2 = 3$, which does not
have capture into the inclination resonances and remains nearly
coplanar throughout its evolution, but we notice that the evolution of
the eccentricities and $\theta_j$ is different from that found by
\citet{lee04} when the outer orbital eccentricity $e_2 \ga 0.45$.
These configurations with $e_2 \ga 0.45$ also do not correspond to any
of the other eccentricity resonance configurations found by
\citet{lee04}.
As we shall see in \S \ref{coplanar}, they belong to a new family of
2:1 eccentricity resonances that can be reached by migration if the
migration rate is faster than that adopted by \citet{lee04} and
$m_1/m_2 \ga 2$.

Gap-opening planets undergo type II migration on the disk viscous
timescale, whose inverse is
\begin{equation}
\left|{\dot a} \over a\right|
\approx {3 \nu \over 2 a^2}
= 9.4 \times 10^{-5} \left(\alpha \over 4\times 10^{-3}\right)
  \left(H/a \over 0.05\right)^2 P^{-1}
\label{vistime}
\end{equation}
\citep{war97}, where $a$ is the semimajor axis of the planet's orbit
about a star of mass $m_0$, ${\dot a} \equiv da/dt$, $\nu = \alpha H^2
\Omega$ is the kinematic viscosity, $\alpha$ is the Shakura-Sunyaev
viscosity parameter, $H$ is the scale height of the disk, and $P =
2\pi/\Omega \approx 2\pi a^{3/2}/(G m_0)^{1/2}$ is the orbital period.
The uncertainties and radial variations in $\alpha$ and $H/a$ mean
that the migration rate can be at least a factor of a few faster or
slower than $10^{-4}/P$.
Although both \citet{tho03} and \citet{lee04} performed three-body
simulations with imposed inward migration on the outer planet only, it
is difficult to determine from the calculations in these papers that
the different results at $e_2 \ga 0.45$ for $m_1/m_2 = 3$ are due to
different migration rates, because they imposed migration in different
ways.
\citet{tho03} adopted $H/a_2 \propto a_2^{1/4}$ so that ${\dot a}_2$
is independent of $a_2$ (with ${\dot a}_2 = -10^{-5}\au\yr^{-1}$ for
most calculations), and they imposed the migration in such a way that
the migration does not slow down after the capture of an inner planet
into resonance.
\citet{lee04} adopted constant $H/a_2$ and performed calculations with
${\dot a}_2/a_2 = -10^{-6}/P_2$ and $-10^{-4}/P_2$, imposed in such a
way that the migration slows down by a factor $\beta/(\beta +
m_1/m_2)$, where $\beta = a_1/a_2 \approx 2^{-2/3}$, after the capture
of an inner planet into 2:1 resonance.
In this paper we examine systematically the effects of different
migration rates (within the range expected for type II migration) using
three-body integrations with migration imposed in the same way.

We also examine systematically the effects of different eccentricity
damping rates during migration.
Significant eccentricity damping can prevent the eccentricities from
reaching high enough values for capture into the new 2:1 eccentricity
resonances or the 4:2 inclination resonances.
There is significant uncertainty in both the sign and the magnitude of
the net effect of planet-disk interaction on the orbital eccentricity
of the planet because of sensitivity to the distribution of disk
material near the locations of the Lindblad and corotation resonances
\citep{gol03,ogi03}.
However, hydrodynamic simulations of two planets orbiting inside an
outer disk have shown eccentricity damping of the outer planet, with
$K = |{\dot e}_2/e_2|/|{\dot a}_2/a_2| \sim 1$ \citep{kle04,kle05}.
\citet{tho03} and \citet{lee04} have reported a small number of
simulations with eccentricity damping for non-coplanar and coplanar
systems, respectively.
In particular, \citet{tho03} have found that the critical value of $K$
for capture into the 4:2 inclination resonances is between 2 and 5 for
$m_1/m_2 = 1$.

In \S \ref{methods} we describe the numerical methods and initial
conditions.
In \S \ref{coplanar} we consider coplanar orbits and show that a
resonant pair can evolve into a new family of 2:1 eccentricity
resonances if the migration rate is faster than that adopted by
\citet{lee04} and $m_1/m_2 \ga 2$, although the new family exists for
$m_1/m_2 \ga 0.2$.
In \S\S \ref{noncoplanar} and \ref{edamping} we consider non-coplanar
orbits.
We show that inclination excitation and capture into the 4:2
inclination resonances are possible for $m_{1}/m_{2} \ga 2$ (as well
as $m_1/m_2 \la 2$), if the migration rate is slower than that adopted
by \citet{tho03}, and that the maximum value of $K = |{\dot e}_2/e_2|/
|{\dot a}_2/a_2|$ for capture into the 4:2 inclination resonances is
of the order of unity.
Our conclusions are summarized and discussed in \S \ref{conclusions}.

\section{NUMERICAL METHODS AND INITIAL CONDITIONS}
\label{methods}

We consider systems consisting of a central star of mass $m_0$, an
inner planet of mass $m_1$, and an outer planet of mass $m_2$, with
$m_1/m_2$ between $0.1$ and $10$.
Unless stated otherwise, $(m_1 + m_2)/m_0 = 10^{-3}$.
For the migration calculations starting with non-resonant orbits, the
planets are initially on circular orbits, with the ratio of the
orbital semimajor axes $\beta = a_1/a_2 = 1/2$ (far from the 2:1
mean-motion commensurability where $\beta \approx 2^{-2/3}$), and the
outer planet is forced to migrate inward.
The calculations presented in \S \ref{coplanar} are for configurations
with exactly coplanar orbits, while those presented in \S\S
\ref{noncoplanar} and \ref{edamping} are for configurations with
initial mutual orbital inclination $i_{\rm mu} = 0.01^\circ$, where
the initial invariable plane is used as the $z = 0$ reference plane.
\citet{tho03} have found that the entry into the inclination
resonances is not strongly influenced by the initial value of
$i_{\rm mu}$, as long as it is $\la 1^\circ$.

The three-body integrations with imposed migration are performed using
the code described in \citet{lee04}, which is a modified version of
the symplectic integrator SyMBA \citep{dun98}.
The outer planet is forced to migrate inward with a migration rate of
the form ${\dot a}_2/a_2 \propto P_2^{-1}$.
The migration slows down by a factor $\beta/(\beta + m_1/m_2)$, where
$\beta \approx 2^{-2/3}$, after the capture of an inner planet into
2:1 resonance (see paragraph with eq. [\ref{vistime}]).
The input and output are in Jacobi orbital elements, and we apply the
forced migration to the Jacobi $a_2$ (and eccentricity damping to the
Jacobi $e_2$ for the calculations with eccentricity damping).
To characterize the new family of 2:1 eccentricity resonances, there
are also calculations in \S \ref{coplanar} with a change in $m_1/m_2$
(and no migration).
The modified SyMBA code used for these calculations is also described
in \citet{lee04}.

\section{A NEW FAMILY OF 2:1 ECCENTRICITY RESONANCES}
\label{coplanar}

\begin{figure}
\epsscale{0.43}
\plotone{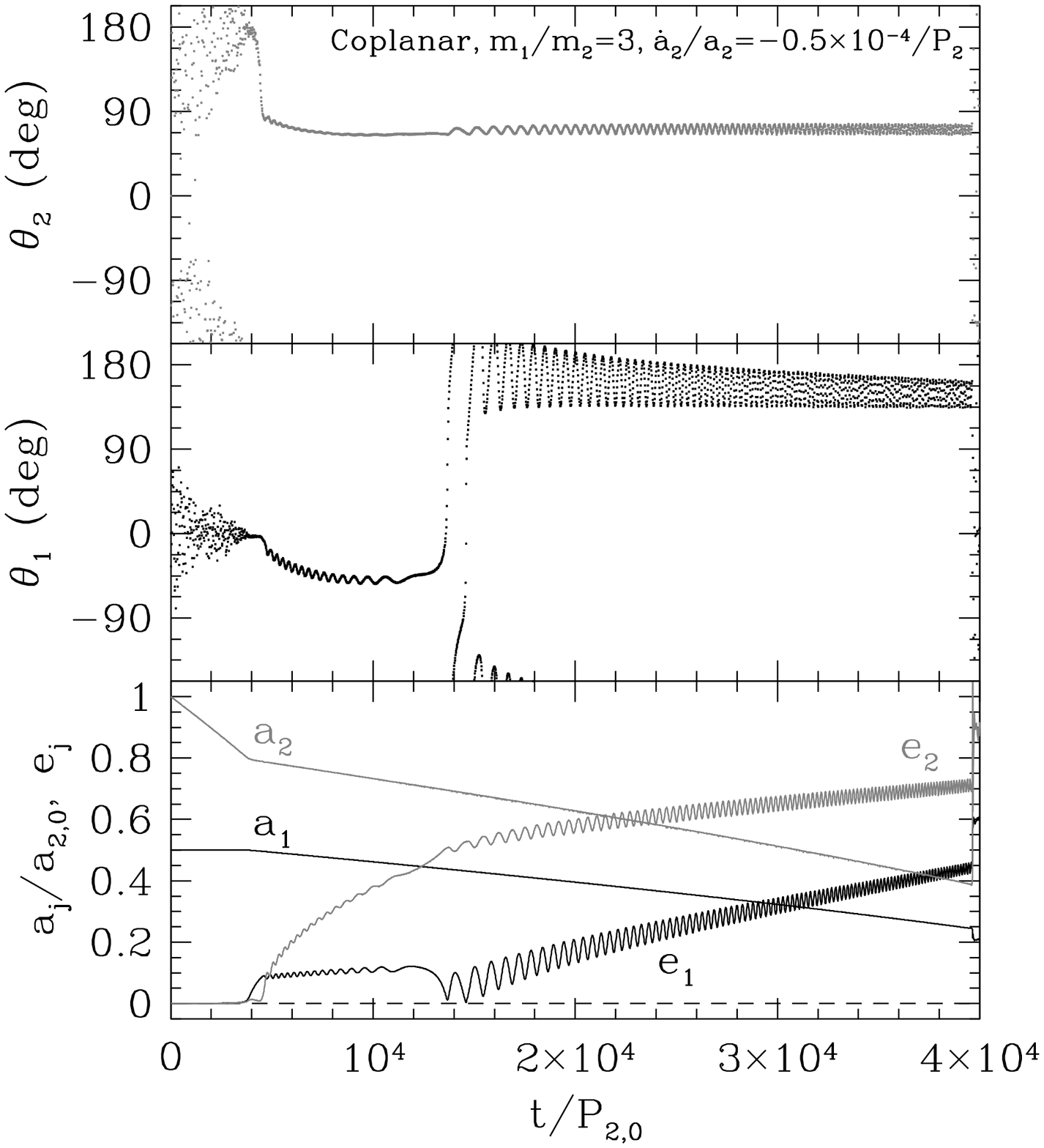}
\caption{\small
Evolution of the semimajor axes $a_1$ and $a_2$, eccentricities $e_1$
and $e_2$, and 2:1 eccentricity-type mean-motion resonance variables
$\theta_1 = \lambda_1 - 2 \lambda_2 + \varpi_1$ and $\theta_2 =
\lambda_1 - 2 \lambda_2 + \varpi_2$ for a differential migration
calculation of coplanar orbits without eccentricity damping.
The mass ratios $(m_1 + m_2)/m_0 = 10^{-3}$ and $m_1/m_2 = 3$.
The outer planet is forced to migrate inward with ${\dot a}_2/a_2 =
-0.5 \times 10^{-4} /P_2$.
The semimajor axes and time are in units of the initial orbital
semimajor axis, $a_{2,0}$, and period, $P_{2,0}$ of the outer planet,
respectively.
The sequence of resonance configurations after resonance capture ---
$(\theta_1, \theta_2) \approx (0^\circ, 180^\circ) \rightarrow$
asymmetric librations --- is identical for $|{\dot a}_2/a_2| \le
10^{-4} /P_2$, but with larger libration amplitudes for faster
migration rate.
\label{fig:coplanarslow}
}
\end{figure}

\begin{figure}
\epsscale{0.43}
\plotone{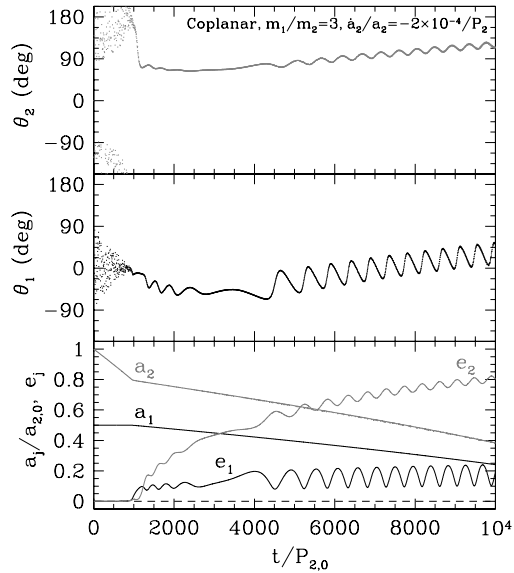}
\caption{\small
Same as Fig. \ref{fig:coplanarslow}, but for the faster migration rate
of ${\dot a}_2/a_2 = -2 \times 10^{-4} /P_2$.
The system enters a new family of 2:1 resonance configurations when
$e_2 \ga 0.45$ (and $t/P_{2,0} \ga 3000$).
\label{fig:coplanarfast}
}
\end{figure}

We begin with migration calculations of coplanar orbits without
eccentricity damping.
We consider $m_1/m_2 = 0.1$, $0.3$, $0.9$, $1.0$, $1.5$, $2.3$,
$2.65$, $3$, $5$, and $10$ (same as in \citealt{lee04}) and
${\dot a}_2/a_2 = -0.5$, $-1$, $-2$, $-4$, and $-8 \times
10^{-4}/P_2$.
Figures \ref{fig:coplanarslow} and \ref{fig:coplanarfast} show the
evolution of the semimajor axes $a_j$, eccentricities $e_j$, and
eccentricity-type resonance variables $\theta_j$ for the calculations
with $(m_1 + m_2)/m_0 = 10^{-3}$, $m_1/m_2 = 3$ and ${\dot a}_2/a_2 =
-0.5 \times 10^{-4} /P_2$ and $-2 \times 10^{-4} /P_2$, respectively.
For ${\dot a}_2/a_2 = -0.5 \times 10^{-4} /P_2$ (Fig.
\ref{fig:coplanarslow}), the sequence of 2:1 resonance
configurations after resonance capture --- from $(\theta_1, \theta_2)
\approx (0^\circ, 180^\circ)$ at small eccentricities to asymmetric
librations of both $\theta_1$ and $\theta_2$ at moderate to large
eccentricities --- is identical to that found by Lee (2004) for
$|{\dot a}_2/a_2| \le 10^{-4} /P_2$, but with larger libration
amplitudes for faster migration rate.
The system eventually becomes unstable at $t/P_{2,0} = 3.96 \times
10^4$, where $P_{2,0}$ is the initial outer orbital period.
For the faster migration rate of ${\dot a}_2/a_2 = -2 \times 10^{-4}
/P_2$ (Fig. \ref{fig:coplanarfast}), the sequence of resonance
configurations is identical to that shown in Figure
\ref{fig:coplanarslow} (but with larger libration amplitudes) when
$e_2 \la 0.45$ (and $t/P_{2,0} \la 3000$).
However, the system enters a new family of 2:1 resonance
configurations when $e_2 \ga 0.45$.
The differences between the configurations in Figures
\ref{fig:coplanarslow} and \ref{fig:coplanarfast} at $e_2 \ga
0.45$ are most obvious in the plots of $e_1$ and $\theta_1$.
We confirm that the configurations in Figure \ref{fig:coplanarfast}
with $e_2 \ga 0.45$ are stable resonance configurations by
integrating the configurations at $t/P_{2,0} = 7000$ and $10^4$
forward with migration turned off and finding stable libration of
$\theta_j$, with the centers and amplitudes of libration nearly
identical to those just before the migration is turned off.\footnote{
Forced migration causes offsets in the libration centers of the
resonance variables \citep{lee04,mur05}.
Both the offsets and the libration amplitudes increase with the
migration rate.
The offsets are typically much smaller than the libration amplitudes
for the asymmetric configurations but could be noticeable for, e.g.,
the $(\theta_1, \theta_2) \approx (0^\circ, 180^\circ)$ configuration
that the system is first captured into (compare
Figs. \ref{fig:coplanarslow} and \ref{fig:coplanarfast}).
}

Migration calculations with different $m_1/m_2$ and ${\dot a}_2/a_2$
show that a system can enter the new family of 2:1 eccentricity
resonances by fast migration if $m_1/m_2 \ga 2$.
For $(m_1 + m_2)/m_0 = 10^{-3}$, the transition occurs between
${\dot a}_2/a_2 = -1 \times 10^{-4} /P_2$ and $-2 \times 10^{-4}/P_2$
for $m_1/m_2 = 2.65$ and $3$, and between ${\dot a}_2/a_2 = -2 \times
10^{-4} /P_2$ and $-4 \times 10^{-4}/P_2$ for $m_1/m_2 = 2.3$, $5$,
and $10$.
However, if ${\dot a}_2/a_2$ is as fast as $-8 \times 10^{-4} /P_2$,
the libration amplitudes are sufficiently large that the system
becomes unstable soon after entering the new family.
Calculations with twice the total planetary mass [$(m_1 + m_2)/m_0 = 2
\times 10^{-3}$] show that the critical migration rate for entry
into the new family is roughly proportional to $(m_1 + m_2)/m_0$.

\begin{figure}
\epsscale{0.45}
\plotone{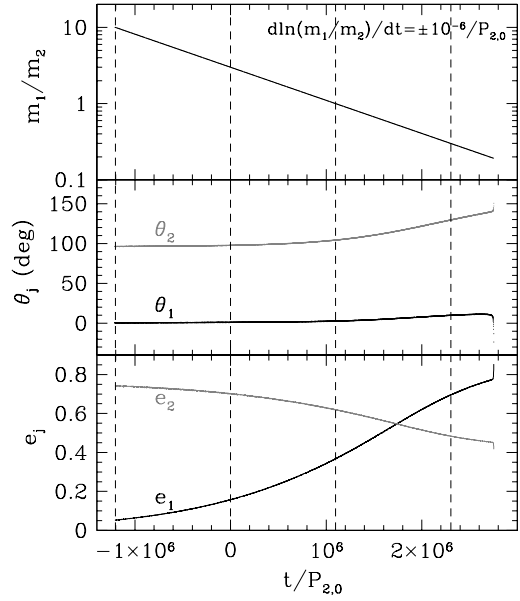}
\caption{\small
Evolution of the eccentricities $e_1$ and $e_2$, eccentricity-type
resonance variables $\theta_1$ and $\theta_2$, and mass ratio
$m_1/m_2$ for calculations in which a configuration in the new family
of 2:1 eccentricity resonances with $m_1/m_2 = 3$ is used as the
starting point and $m_1/m_2$ is increased and decreased.
The starting configuration with small libration amplitudes is obtained
by adjusting the orbital parameters of the large-libration-amplitude
configuration at $t/P_{2,0} = 7000$ in Fig. \ref{fig:coplanarfast}.
The results from the calculations with $d \ln(m_1/m_2)/dt =
-10^{-6}/P_{2,0}$ and $10^{-6}/P_{2,0}$ are plotted along the positive
and negative time axis, respectively.
The system becomes unstable when $m_1/m_2$ is decreased to about
$0.2$.
The configurations with $m_1/m_2 = 0.3$, $1$, $3$, and $10$, indicated
by the dashed lines, are used as initial conditions for calculations
in Fig. \ref{fig:loci}{\it a}.
\label{fig:masschange}
}
\end{figure}

\begin{figure}
\epsscale{1.0}
\plotone{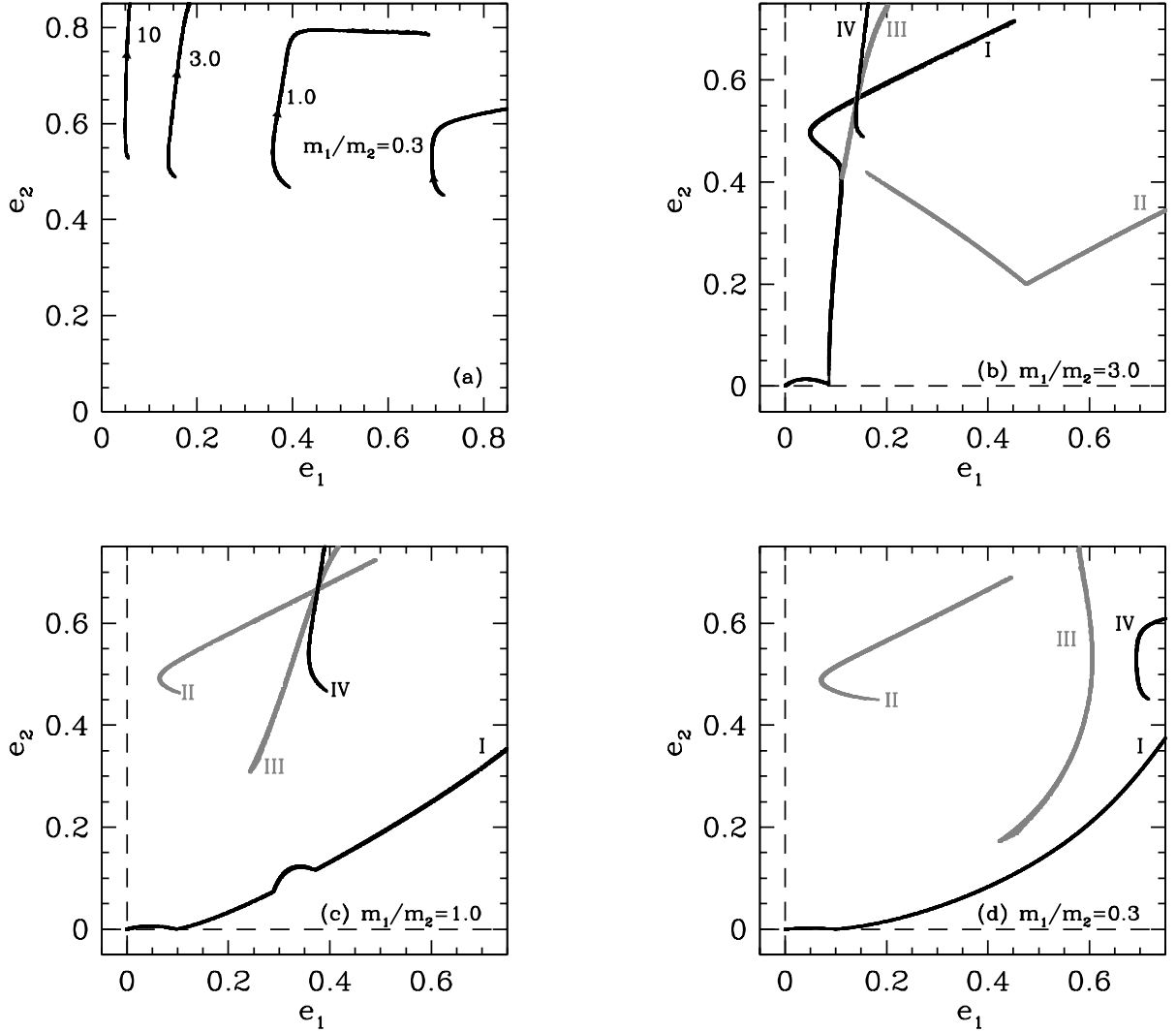}
\caption{\small
Loci in the $e_1$-$e_2$ plane of coplanar 2:1 resonance
configurations.
({\it a}) Configurations in the new family of eccentricity resonances
from inward and outward migration calculations with initial
conditions ({\it triangles}, oriented to indicate the direction for
inward migration) from Fig. \ref{fig:masschange} for $m_1/m_2 = 0.3$,
$1$, $3$, and $10$.
Comparison of the new family (labeled IV) with the families (labeled
I--III) found by \citet{lee04} for $m_1/m_2 =$ ({\it b}) $3$,
({\it c}) $1$, and ({\it d}) $0.3$, respectively.
It is possible to enter the new family by fast migration only for
$m_1/m_2 \ga 2$ because the combination that sequences I and IV come
close to each other and that some configurations along sequence I have
asymmetric librations and intersecting orbits occurs only for $m_1/m_2
\ga 2$ (see text for details).
\label{fig:loci}
}
\end{figure}

To find the small-libration-amplitude (or near exact resonance)
counterpart for this new family and to determine the range of
$m_1/m_2$ for which this family exists, we take the
large-libration-amplitude configuration at $t/P_{2,0} = 7000$ in
Figure \ref{fig:coplanarfast} and adjust the orbital parameters to
obtain a small-libration-amplitude configuration with $m_1/m_2 =3$,
$e_1 = 0.158$, $e_2 = 0.702$, $\theta_1 = 1^\circ$, and $\theta_2 =
98^\circ$.
This small-libration-amplitude configuration is used as the starting
point for two calculations in which $m_1/m_2$ is increased or
decreased slowly to find a sequence of configurations with different
$m_1/m_2$.
The results are shown in Figure \ref{fig:masschange}, with the
calculations with $d \ln(m_1/m_2)/dt = -10^{-6}/P_{2,0}$ and
$10^{-6}/P_{2,0}$ along the positive and negative time axis,
respectively.
The inner eccentricity $e_1$ increases (and outer eccentricity $e_2$
decreases) with decreasing $m_1/m_2$, and the system becomes unstable
when $m_1/m_2$ is decreased to about $0.2$.
The resonance configuration for a given $m_1/m_2$ from Figure
\ref{fig:masschange} is then used as the starting point for slow
inward (${\dot a}_2/a_2 = -10^{-6} /P_2$) and outward (${\dot a}_2/a_2
= 10^{-6} /P_2$) migration calculations to search for other resonance
configurations with the same $m_1/m_2$.
(The slow rate of change in $m_1/m_2$ or $a_2$ in these calculations
ensures that the libration amplitudes and offsets remain small.)
Figure \ref{fig:loci}{\it a} shows the loci in the $e_1$-$e_2$ plane
of the stable resonance configurations from the migration calculations
with $m_1/m_2 = 0.3$, $1$, $3$, and $10$.
The initial conditions (dashed lines in Fig. \ref{fig:masschange}) are
indicated by the triangles in Figure \ref{fig:loci}{\it a}, and the
results from inward (outward) migration extend above (below) the
triangles.
In addition to the calculations shown in Figure \ref{fig:masschange},
we perform several calculations in which different configurations
along the locus shown in Figure \ref{fig:loci}{\it a} for $m_1/m_2 =
0.3$ are used as the starting point and $m_1/m_2$ is decreased.
In all cases, the system becomes unstable when $m_1/m_2$ is decreased
to about $0.2$.
Thus the new family of 2:1 eccentricity resonances does not appear to
exist for $m_1/m_2 \la 0.2$.

The new family of 2:1 eccentricity resonances has $e_2 \ga
0.4$--$0.5$, asymmetric librations, and intersecting orbits, and it is
distinct from any of the families found by \citet{lee04}.
In Figures \ref{fig:loci}{\it b}--{\it d}, we compare the new family
(labeled IV) with the families (labeled I--III) found by \citet{lee04}
for $m_1/m_2 = 3$, $1$, and $0.3$, respectively.
Sequence I is the sequence reached by slow migration of planets with
constant masses and initially nearly circular orbits;
sequence II was found by a combination of calculations in which
$m_1/m_2$ is changed and slow migration calculations;
and sequence III consists of configurations with $(\theta_1, \theta_2)
\approx (180^\circ, 0^\circ)$.
For $m_1/m_2 = 3$ (Fig. \ref{fig:loci}{\it b}), sequences I and IV
come close to each other.
In addition, like sequence IV, the configurations in sequence I with
$e_2 \ga 0.34$ have intersecting orbits, as well as asymmetric
librations \citep{lee04}.
Thus it is possible to jump from sequence I to sequence IV if the
libration amplitudes are large due to fast migration, as shown in
Figure \ref{fig:coplanarfast}.
For $m_1/m_2 = 1$ (Fig. \ref{fig:loci}{\it c}), sequences I and IV do
not come close to each other.
For $m_1/m_2 = 0.3$ (Fig. \ref{fig:loci}{\it d}), although sequences I
and IV come relatively close to each other at large $e_1$, the
configurations in sequence I at large $e_1$ have $(\theta_1, \theta_2)
\approx (0^\circ, 0^\circ)$ and non-intersecting orbits, and it is not
possible to jump to sequence IV with asymmetric librations and
intersecting orbits.
Because the combination that sequences I and IV come close to each
other and that some configurations along sequence I have asymmetric
librations and intersecting orbits occurs only for $m_1/m_2 \ga 2$, we
can understand why it is possible to enter the new family by fast
migration only for $m_1/m_2 \ga 2$, even though the new family exists
for $m_1/m_2 \ga 0.2$.

The existence of the new family of 2:1 eccentricity resonances was
noted by \citet{lt04}.
\citet{voy05} have also discovered this family in their
search for both stable and unstable asymmetric periodic orbits at the
2:1 resonance for three cases with $m_1/m_2 = 0.54$, $1$, and $1.86$.
The stable periodic orbits are resonance configurations with zero
libration amplitudes.
In the range of $m_1/m_2$ studied by \citet{voy05} (i.e., $m_1/m_2 \la
2.75$, \citealt{lee04}), sequence II has loci similar to those shown
in Figures \ref{fig:loci}{\it c} and \ref{fig:loci}{\it d}, and
\citet{voy05} found that sequences II and IV (the new family) are
connected to each other by a sequence of {\it unstable} periodic
orbits.

\section{THE 4:2 INCLINATION RESONANCES}
\label{noncoplanar}

\begin{figure}
\epsscale{1.0}
\plottwo{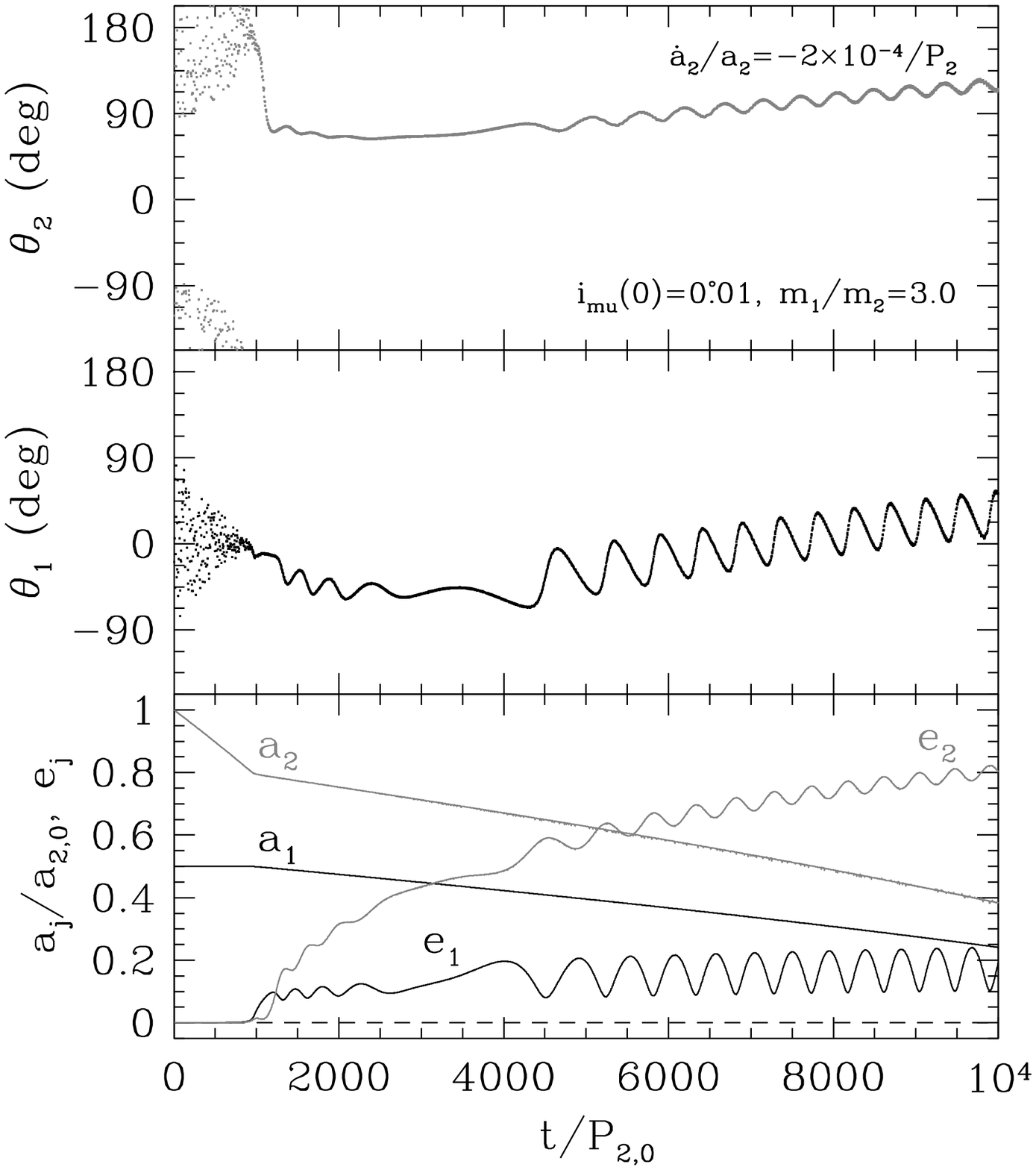}{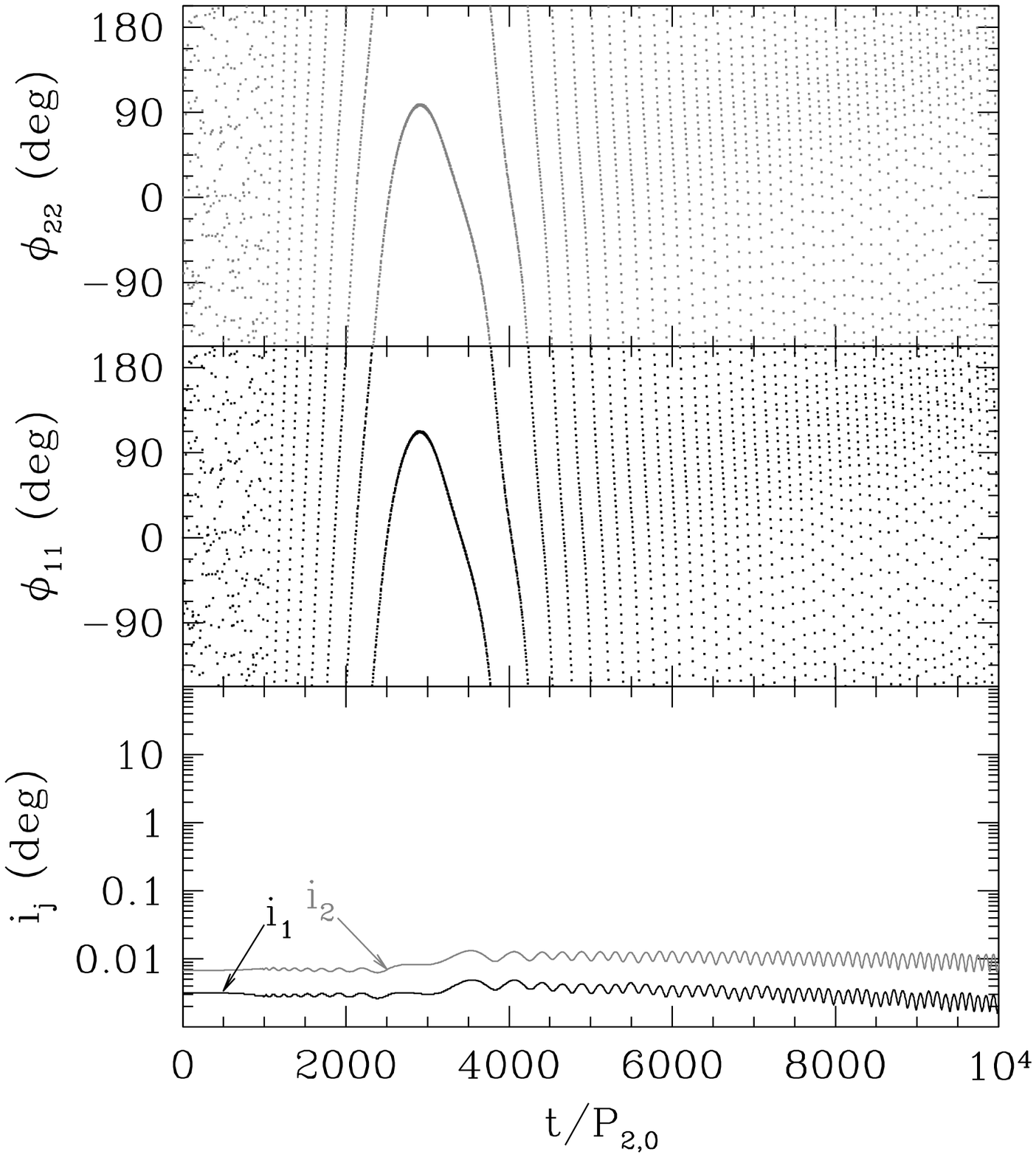}
\caption{\small
Evolution of the semimajor axes $a_1$ and $a_2$, eccentricities $e_1$
and $e_2$, 2:1 eccentricity-type resonance variables $\theta_1$ and
$\theta_2$, inclinations $i_1$ and $i_2$, and 4:2 inclination-type
resonance variables $\phi_{11} = 2 \lambda_1 - 4 \lambda_2 + 2
\Omega_1$ and $\phi_{22} = 2 \lambda_1 - 4 \lambda_2 + 2 \Omega_2$ for
a differential migration calculation of non-coplanar orbits without
eccentricity damping.
The mass ratios $(m_1 + m_2)/m_0 = 10^{-3}$ and $m_1/m_2 = 3$, and the
initial mutual orbital inclination $i_{\rm mu} = 0.01^\circ$.
The outer planet is forced to migrate inward with the fast migration
rate of ${\dot a}_2/a_2 = -2 \times 10^{-4} /P_2$.
The eccentricity resonances enter the new family as in the planar case
shown in Fig. \ref{fig:coplanarfast}, and there is no capture into the
inclination resonances or excitation of the inclinations.
\label{fig:inclinedfast}
}
\end{figure}

\begin{figure}
\epsscale{1.0}
\plottwo{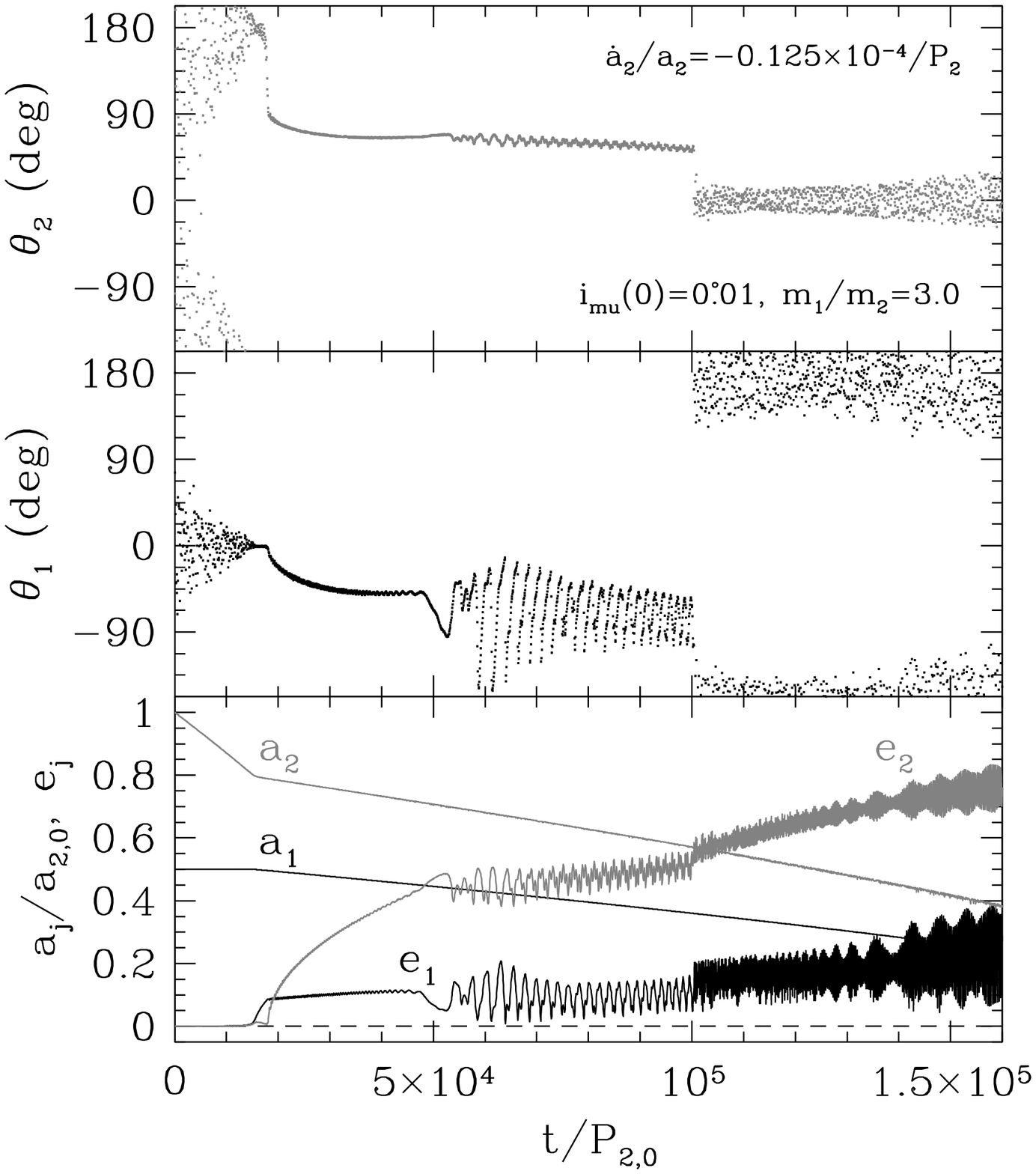}{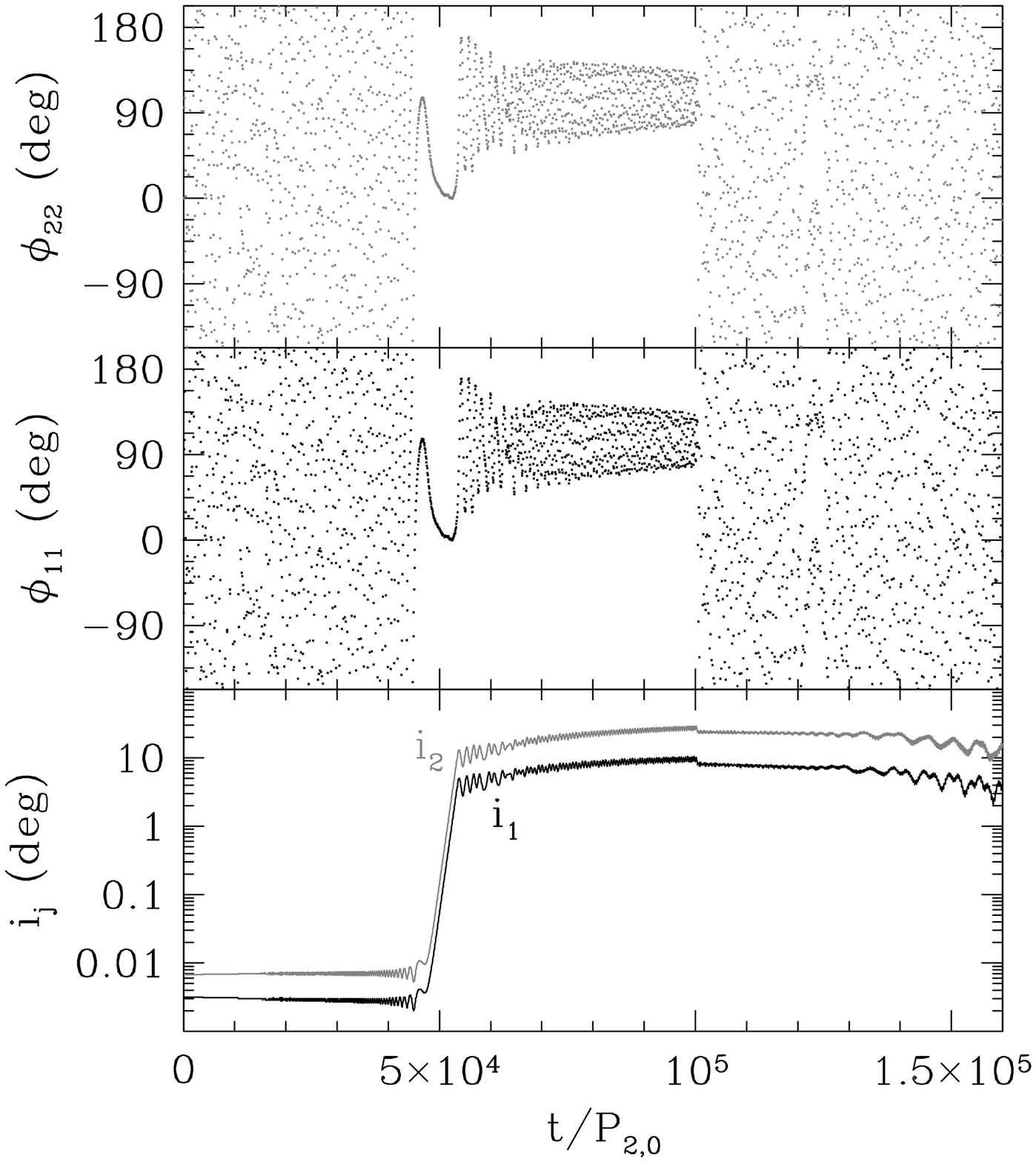}
\caption{\small
Same as Fig. \ref{fig:inclinedfast}, but for the slow migration rate
of ${\dot a}_2/a_2 = -0.125 \times 10^{-4} /P_2$.
The system is initially captured into the 2:1 eccentricity resonances
only.
There is a phase from $t/P_{2,0} \approx 4.7 \times 10^4$ to
$5.3 \times 10^4$ with $\phi_{jj}$ changing slowly and $i_j$
increasing rapidly before $\phi_{jj}$ are captured into libration.
The system eventually evolves out of the inclination resonances at
$t/P_{2,0} \approx 1.0 \times 10^5$, and the eccentricity resonances
switch to the $\theta_1 \approx 180^\circ$ and $\theta_2 \approx
0^\circ$ configuration.
\label{fig:inclinedslow}
}
\end{figure}

\begin{figure}
\epsscale{1.0}
\plottwo{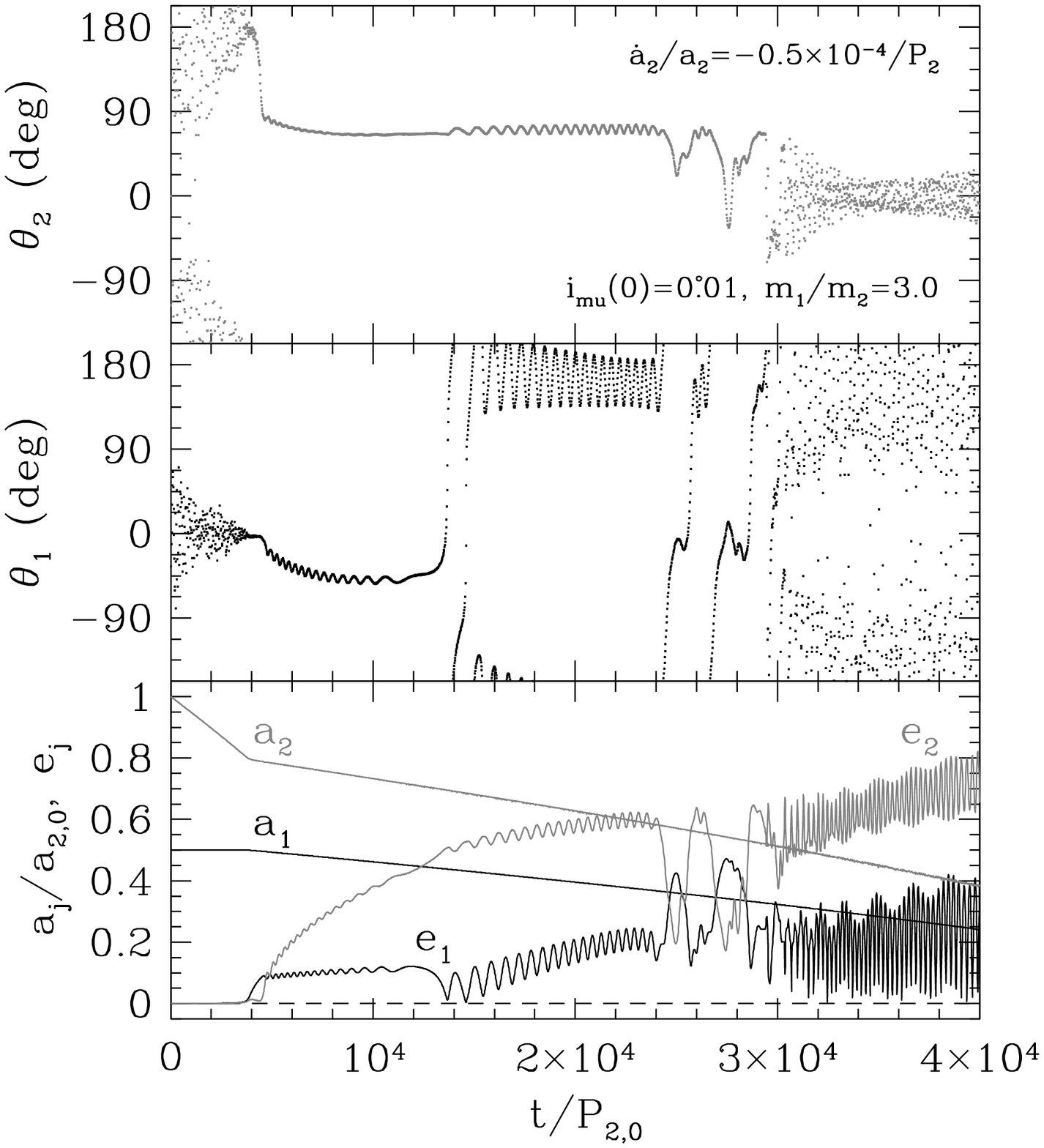}{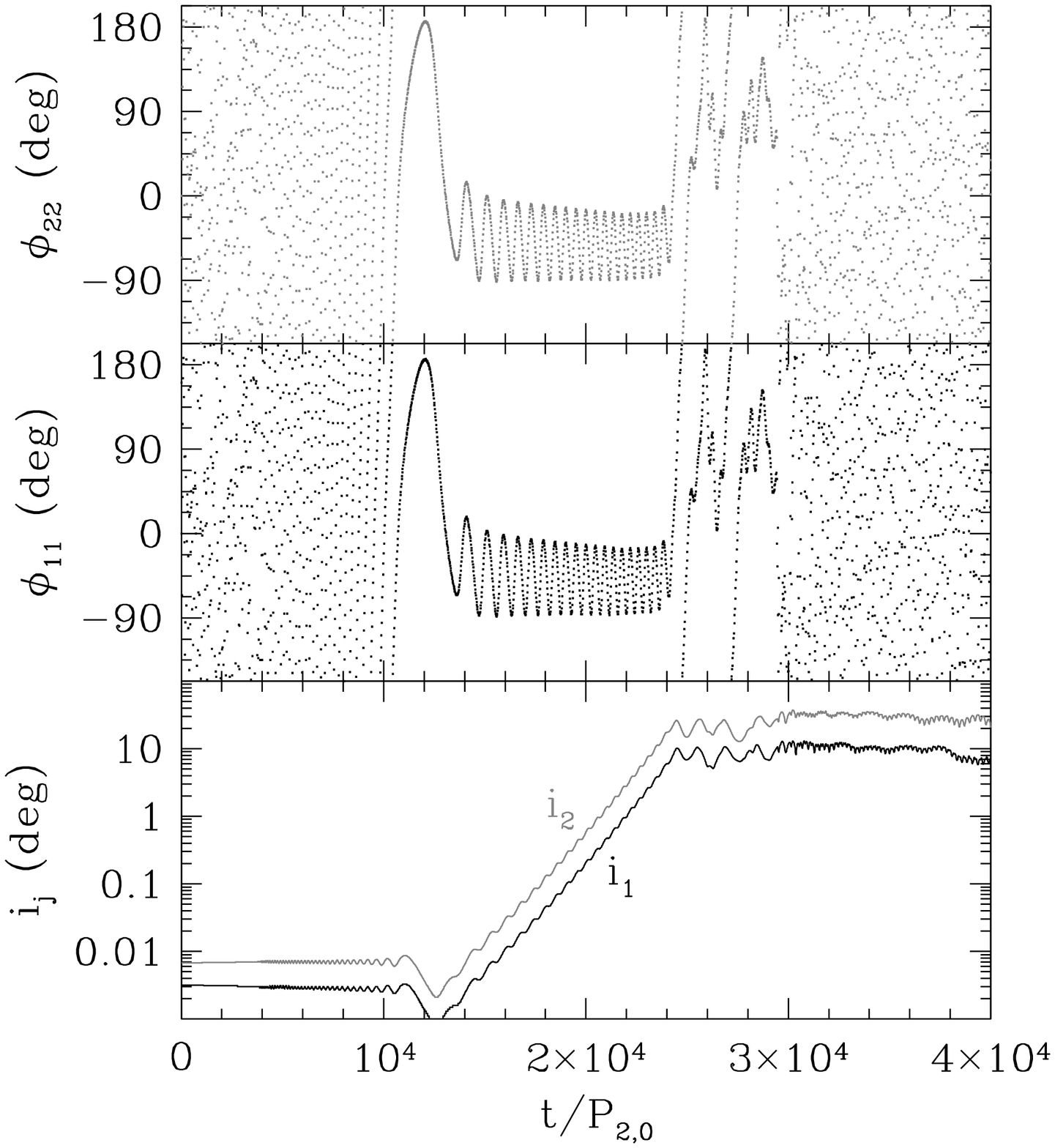}
\caption{\small
Same as Fig. \ref{fig:inclinedfast}, but for the intermediate
migration rate of ${\dot a}_2/a_2 = -0.5 \times 10^{-4} /P_2$.
The rapid inclination excitation phase occurs from $t/P_{2,0} \approx
1.2 \times 10^4$ to $2.4 \times 10^4$.
Then $\phi_{jj}$ and $\theta_1$ alternate between libration and
circulation for about $6000 P_{2,0}$, before $\phi_{jj}$ change to
circulation only and the eccentricity resonances to the $\theta_1
\approx 180^\circ$ and $\theta_2 \approx 0^\circ$ configuration.
\label{fig:inclinedmed}
}
\end{figure}

We consider next migration calculations of non-coplanar orbits without
eccentricity damping.
We perform calculations with initial mutual orbital inclination
$i_{\rm mu} = 0.01^\circ$ and migration rate ${\dot a}_2/a_2 =
-0.125$, $-0.25$, $\ldots$, $-4 \times 10^{-4}/P_2$.
Figures \ref{fig:inclinedfast}, \ref{fig:inclinedslow}, and
\ref{fig:inclinedmed} show the results for $m_1/m_2 = 3$ and
${\dot a}_2/a_2 = -2$, $-0.125$, and $-0.5 \times 10^{-4}/P_2$,
respectively.
In each figure, we plot the evolution of the inclinations $i_j$ and
the inclination-type resonance variables $\phi_{jj}$, as well as
$a_j$, $e_j$, and $\theta_j$.
The evolution of $\phi_{11}$ is nearly identical to that of
$\phi_{22}$, which means that $\Omega_1 - \Omega_2 = (\phi_{11} -
\phi_{22})/2$ (which is $180^\circ$ initially due to our choice of the
initial invariable plane as the $z = 0$ reference plane) is close to
$180^\circ$ throughout and the ascending nodes are nearly antialigned.

The fast migration calculation shown in Figure \ref{fig:inclinedfast}
is similar to that in Figure \ref{fig:coplanarfast}, but with
non-coplanar orbits.
The $e_j$ and $\theta_j$ evolve as in the planar case shown in
Figure \ref{fig:coplanarfast} and enter the new family of eccentricity
resonances when $e_2 \ga 0.45$.
There is no capture into the inclination resonances or excitation of
the inclinations.
Note, however, that the circulation of the inclination resonance
variables $\phi_{jj}$ changes from prograde to retrograde at about the
same time as the entry into the new family of eccentricity resonances.
The evolution in this figure is that found by \citet{tho03} in their
simulation with $m_1/m_2 = 3$.

As in the planar case, the non-coplanar calculations with $m_1/m_2 =
3$ and ${\dot a}_2/a_2$ slower than $-2 \times 10^{-4} /P_2$ do not
enter the new family of eccentricity resonances.
However, unlike the planar case, the evolution for ${\dot a}_2/a_2$
slower than $-0.5 \times 10^{-4} /P_2$ is qualitatively different
from that for ${\dot a}_2/a_2 = -0.5$ and $-1 \times 10^{-4} /P_2$.
Figure \ref{fig:inclinedslow} shows a calculation with a slow
migration rate (${\dot a}_2/a_2 = -0.125 \times 10^{-4}/P_2$).
The system is initially captured into the 2:1 eccentricity resonances
only, and the initial evolution after capture is similar to that in
the planar calculation with slow migration shown in Figure
\ref{fig:coplanarslow}.
But starting at $t/P_{2,0} \approx 4.7 \times 10^4$ (when $e_2 \approx
0.45$), the inclination resonance variables $\phi_{jj}$ change very
slowly for about $6000 P_{2,0}$, and the inclinations increase
rapidly.
It is likely that this slow change of $\phi_{jj}$ is associated with
the proximity to the separatrix of the inclination resonances, since
the circulation/libration period is infinite on the separatrix.
We can understand qualitatively the almost exponential growth in the
inclinations by noting that the lowest order inclination resonance
terms at the 2:1 commensurability in the disturbing potential $\Phi$
are second order and proportional to $i_1^2 \cos\phi_{11}$, $i_1 i_2
\cos\phi_{12}$, and $i_2^2 \cos\phi_{22}$.
Thus the lowest order terms for $di_j/dt \propto i_j^{-1}
{\partial \Phi / \partial \Omega_j}$ are proportional to $i_j
\sin\phi_{jj}$ and $i_k \sin\phi_{12}$ (where $k = 2$ for $j = 1$ and
vice versa), which can result in exponential growth if $\phi_{jj}$ and
$\phi_{12}$ are not equal to $0^\circ$ or $180^\circ$ and change very
slowly.
At $t/P_{2,0} \approx 5.3 \times 10^4$, both $\phi_{11}$ and
$\phi_{22}$ are captured into resonance and librate about $110^\circ$,
and the inclinations increase slowly due to the continued migration
forcing the system deeper into inclination resonances.
We note that the simultaneous librations of $\theta_j$ and $\phi_{jj}$
affect the values of $\theta_j$ and $e_j$ during this phase (compare
Figs. \ref{fig:coplanarslow} and \ref{fig:inclinedslow}).
As in the case of the eccentricity resonances, asymmetric libration of
$\phi_{jj}$ about an angle other than $0^\circ$ or $180^\circ$ is
possible when the inclinations are not small and $di_j/dt$ is not
dominated by the lowest order terms in the disturbing potential.\footnote{
In the limit of the circular, planar, restricted, three-body problem,
one can identify the terms that give rise to asymmetric libration for
the $n$:1 (not just 2:1) exterior resonance as coming from the
indirect part of the disturbing potential, and there is a qualitative
physical explanation based on the indirect acceleration imparted on
the test particle over a synodic period \citep{pan04,mur05}.
This type of analysis has not been generalized to either the planar
two-planet problem with two resonance variables $\theta_1$ and
$\theta_2$ or the inclination resonances.
}
We confirm that the configuration at, e.g., $t/P_{2,0} = 8.0 \times
10^4$ in Figure \ref{fig:inclinedslow} is indeed in stable inclination
resonances by taking that configuration as the starting point for a
three-body integration {\it without} forced migration and finding
stable libration of $\phi_{jj}$ about $110^\circ$ and no secular
change in $i_j$ throughout that integration.
In Figure \ref{fig:inclinedslow} the system eventually evolves out of
the inclination resonances at $t/P_{2,0} \approx 1.0 \times 10^5$, and
the eccentricity resonances switch to the $\theta_1 \approx 180^\circ$
and $\theta_2 \approx 0^\circ$ configuration.
As mentioned in \S \ref{intro}, \cite{tho03} have seen similar
switching to the $\theta_1 \approx 180^\circ$ and $\theta_2 \approx
0^\circ$ configuration in their simulations with $m_1/m_2 = 1$.

\begin{figure}
\epsscale{1.0}
\plottwo{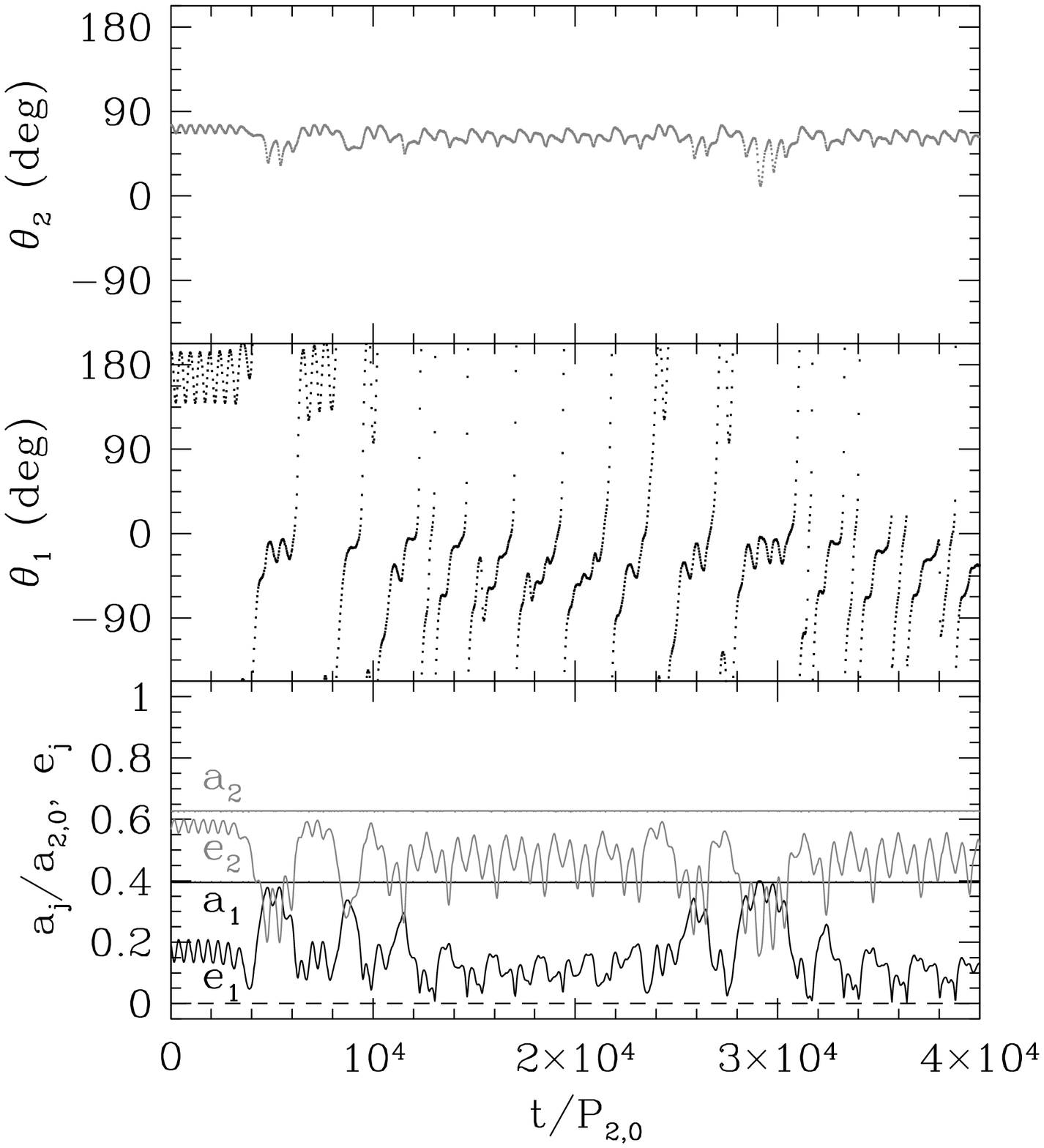}{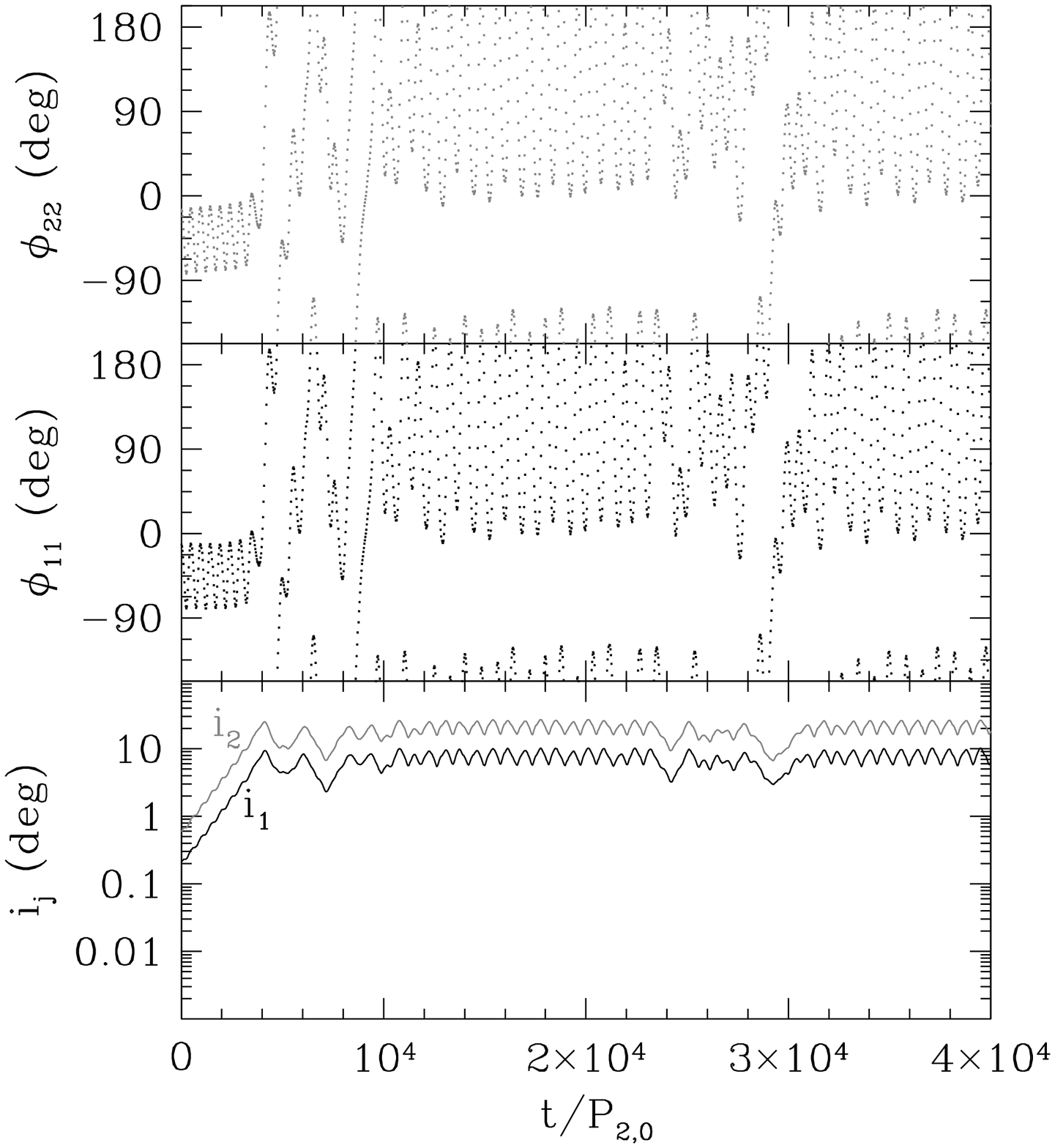}
\caption{\small
Evolution for the calculation in which the configuration at $t/P_{2,0}
= 2.0 \times 10^4$ in Fig. \ref{fig:inclinedmed} is used as the
starting point for a three-body integration without forced migration.
The inclinations continue to increase rapidly for about $4000
P_{2,0}$, and the evolution of all the plotted variables for the first
$10^4 P_{2,0}$ is similar to that between $t/P_{2,0} = 2.0 \times
10^4$ and $3.0 \times 10^4$ in Fig. \ref{fig:inclinedmed} with
migration.
\label{fig:nomigration}
}
\end{figure}

In contrast to the overall evolution on the migration timescale, the
time spent in the phase with $\phi_{jj}$ changing slowly and $i_j$
increasing rapidly is nearly independent of the migration rate for
slow migration.
Thus this phase takes up a larger and larger fraction of the total
evolution time with increasing migration rate, and there is no longer
a phase with $\phi_{jj}$ clearly in resonance if the migration rate
${\dot a}_2/a_2$ is as fast as $-0.5$ and $-1 \times 10^{-4} /P_2$.
(Even faster migration rate would result in entry into the new family
of eccentricity resonances and no inclination excitation, as discussed
above.)
For ${\dot a}_2/a_2 = -0.5 \times 10^{-4} /P_2$
(Fig. \ref{fig:inclinedmed}), the rapid inclination excitation phase
occurs from $t/P_{2,0} \approx 1.2 \times 10^4$ to $2.4 \times 10^4$.
Then $\phi_{jj}$, as well as $\theta_1$, alternate between libration
and circulation for about $6000 P_{2,0}$, before $\phi_{jj}$ change to
circulation only and the eccentricity resonances to the $\theta_1
\approx 180^\circ$ and $\theta_2 \approx 0^\circ$ configuration, with
$\theta_1$ nearly circulating but spending most of its time around
$180^\circ$.
The oscillations of the inclination resonance variables $\phi_{jj}$
between $t/P_{2,0} \approx 1.2 \times 10^4$ to $2.4 \times 10^4$ in
Figure \ref{fig:inclinedmed} might lead one to think that $\phi_{jj}$
are in resonance and librating about equilibrium values and that the
rapid increase in the inclinations is due to continued migration
forcing the system deeper into inclination resonances.
However, this would be inconsistent with our earlier observation that
the duration of this phase is nearly independent of the migration rate
for slow migration.
To show that this rapid inclination excitation is in fact {\it not}
due to migration forcing, we take the configuration at $t/P_{2,0} =
2.0 \times 10^4$ in Figure \ref{fig:inclinedmed} as the starting point
for a three-body integration without forced migration.
The results are shown in Figure \ref{fig:nomigration}.
As we can see, the inclinations continue to increase rapidly for about
$4000 P_{2,0}$ even without forced migration.
Furthermore, the evolution of all the plotted variables for the first
$10^4 P_{2,0}$ in Figure \ref{fig:nomigration} without migration is
similar to that between $t/P_{2,0} = 2.0 \times 10^4$ and $3.0 \times
10^4$ in Figure \ref{fig:inclinedmed} with migration.
Figure \ref{fig:nomigration} also shows us what would happen if the
migration stops due to, e.g., disk dispersal when the system is in the
phase with rapid inclination excitation.
The inclinations would continue to increase for a while, and the
inclination resonance variables would eventually end up in
large-amplitude libration (alternating with circulation to varying
degree).

\begin{figure}
\epsscale{0.45}
\plotone{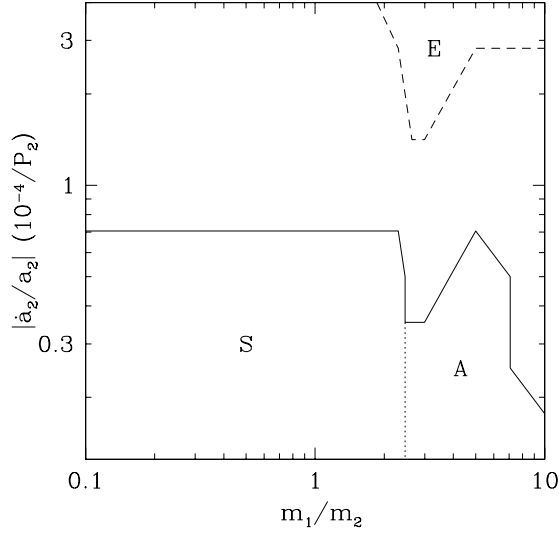}
\caption{\small
Types of evolution for different $m_1/m_2$ and ${\dot a}_2/a_2$.
The results are from migration calculations with $(m_1 + m_2)/m_0 =
10^{-3}$, non-coplanar orbits, and no eccentricity damping.
In the region labeled E, the eccentricity resonances enter the new
family, and there is no capture into inclination resonances or
excitation of the inclinations.
In the region below the solid line, the inclination resonance
variables $\phi_{jj}$ are captured into libration (symmetric in the
region labeled S and asymmetric in the region labeled A) after a phase
with $\phi_{jj}$ changing slowly and $i_j$ increasing rapidly.
In the unlabeled region, there is typically a rapid inclination
excitation phase, but not a phase with $\phi_{jj}$ clearly in
resonance.
\label{fig:summary}
}
\end{figure}

Figure \ref{fig:summary} summarizes the results for $(m_1 + m_2)/m_0 =
10^{-3}$ and different $m_1/m_2$ and ${\dot a}_2/a_2$.
For $m_1/m_2 \ga 2$ and fast migration (the region labeled E in
Fig. \ref{fig:summary}), the eccentricity resonances enter the new
family, and there is no capture into inclination resonances or
excitation of the inclinations.
For slow migration (the region below the solid line in
Fig. \ref{fig:summary}), the inclination resonance variables
$\phi_{jj}$ are captured into libration after a phase with $\phi_{jj}$
changing slowly and $i_j$ increasing rapidly.
The inclination resonance configuration is symmetric with $\phi_{jj}$
librating about $180^\circ$ in the region labeled S with $m_1/m_2 \la
2.5$ (see \citealt{tho03} for an example with $m_1/m_2 = 1$), and it
is asymmetric with $\phi_{jj}$ librating about an angle other than
$0^\circ$ or $180^\circ$ in the region labeled A with $m_1/m_2 \ga
2.5$ (e.g., Fig. \ref{fig:inclinedslow}).
For intermediate migration rate (and also fast migration rate if
$m_1/m_2 \la 2$), we typically see the rapid inclination excitation
phase, but not a phase with $\phi_{jj}$ clearly in resonance (e.g.,
Fig. \ref{fig:inclinedmed}).
The inclination excitation can be partial, with the mutual inclination
reaching a maximum of $\sim 1^\circ$ or less, if $m_1/m_2$ is large
(in particular $m_1/m_2 = 10$).

In the phase with simultaneous librations of the eccentricity and
inclination resonance variables, we can see from the definitions of
$\theta_j$ (eq. [\ref{theta1}]--[\ref{theta2}]) and $\phi_{jj}$
(eq. [\ref{phi11}]--[\ref{phi22}]) that the arguments of periapse
$\omega_j = \varpi_j - \Omega_j = \theta_j - \phi_{jj}/2$ also
librate.
For $m_1/m_2 \la 2.5$ with symmetric libration of $\phi_{jj}$,
$\omega_j$ librate about $\pm 90^\circ$ (i.e., the periapse is on
average $90^\circ$ ahead of or behind the ascending node), while for
$m_1/m_2 \ga 2.5$ with asymmetric libration of $\phi_{jj}$, the
libration of $\omega_j$ is also asymmetric.

\section{EFFECTS OF ECCENTRICITY DAMPING}
\label{edamping}

As we mentioned in \S \ref{intro}, sufficient eccentricity damping can
prevent the eccentricities from reaching high enough values for
inclination excitation and/or capture into the inclination resonances.
In order to study the effects of eccentricity damping, we repeat the
non-coplanar calculations in \S \ref{noncoplanar} with the ratio of
eccentricity damping to migration of the outer planet, $K =
|{\dot e}_2/e_2|/|{\dot a}_2/a_2|$, ranging from $0.25$ to $8$.

\begin{figure}
\epsscale{0.45}
\plotone{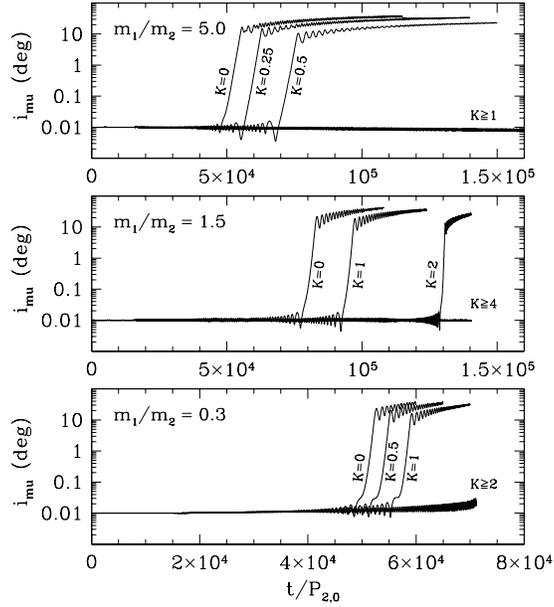}
\caption{\small
Evolution of the mutual inclination $i_{\rm mu}$ for $m_1/m_2 = 0.3$,
$1.5$, and $5.0$, ${\dot a}_2/a_2 = -0.125 \times 10^{-4} /P_2$, and
different eccentricity damping ratio $K = |{\dot e}_2/e_2|/
|{\dot a}_2/a_2|$.
The eccentricities never reach high enough values for inclination
excitation and capture into inclination resonances when $K$ exceeds a
critical value.
\label{fig:edampslow}
}
\end{figure}

\begin{figure}
\epsscale{0.45}
\plotone{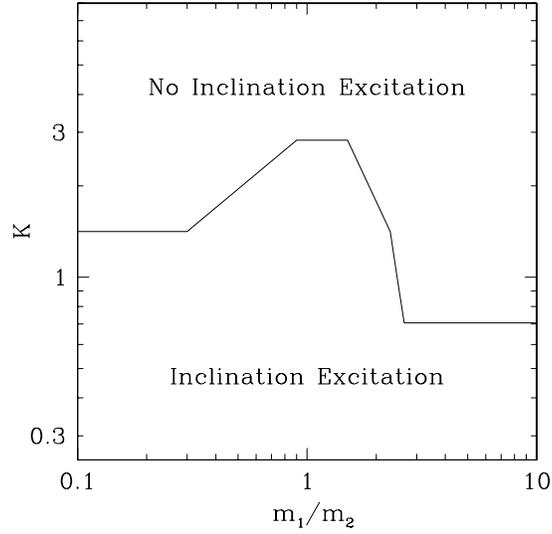}
\caption{\small
Critical value of $K$ as a function of $m_1/m_2$.
The critical value is for capture into inclination resonances for slow
migration (${\dot a}_2/a_2$ below the solid line in
Fig. \ref{fig:summary}) and for inclination excitation (which may or
may not be followed by a phase with $\phi_{jj}$ clearly in resonance)
for migration rate up to ${\dot a}_2/a_2 = -2 \times 10^{-4} /P_2$.
\label{fig:edampsummary}
}
\end{figure}

We consider first the calculations with slow migration
(${\dot a}_2/a_2$ below the solid line in Fig. \ref{fig:summary}).
Figure \ref{fig:edampslow} shows the evolution of the mutual
inclination $i_{\rm mu}$ for $m_1/m_2 = 0.3$, $1.5$, and $5.0$,
${\dot a}_2/a_2 = -0.125 \times 10^{-4} /P_2$, and different $K$.
As $K$ increases from zero, the system enters the rapid inclination
excitation phase and the subsequent capture into the 4:2 inclination
resonances later and later, because the eccentricities grow slower and
slower.
However, when $K$ exceeds a critical value, the eccentricities
never reach high enough values for inclination excitation and capture
into inclination resonances.
The critical value of $K$ is $\approx 1.4$ for $m_1/m_2 \la 0.3$,
$\approx 2.8$ for $m_1/m_2 \approx 0.9$--$1.5$, and $\approx 0.7$ for
$m_1/m_2 \ga 2.65$ (Fig. \ref{fig:edampsummary}).

\begin{figure}
\epsscale{1.0}
\plottwo{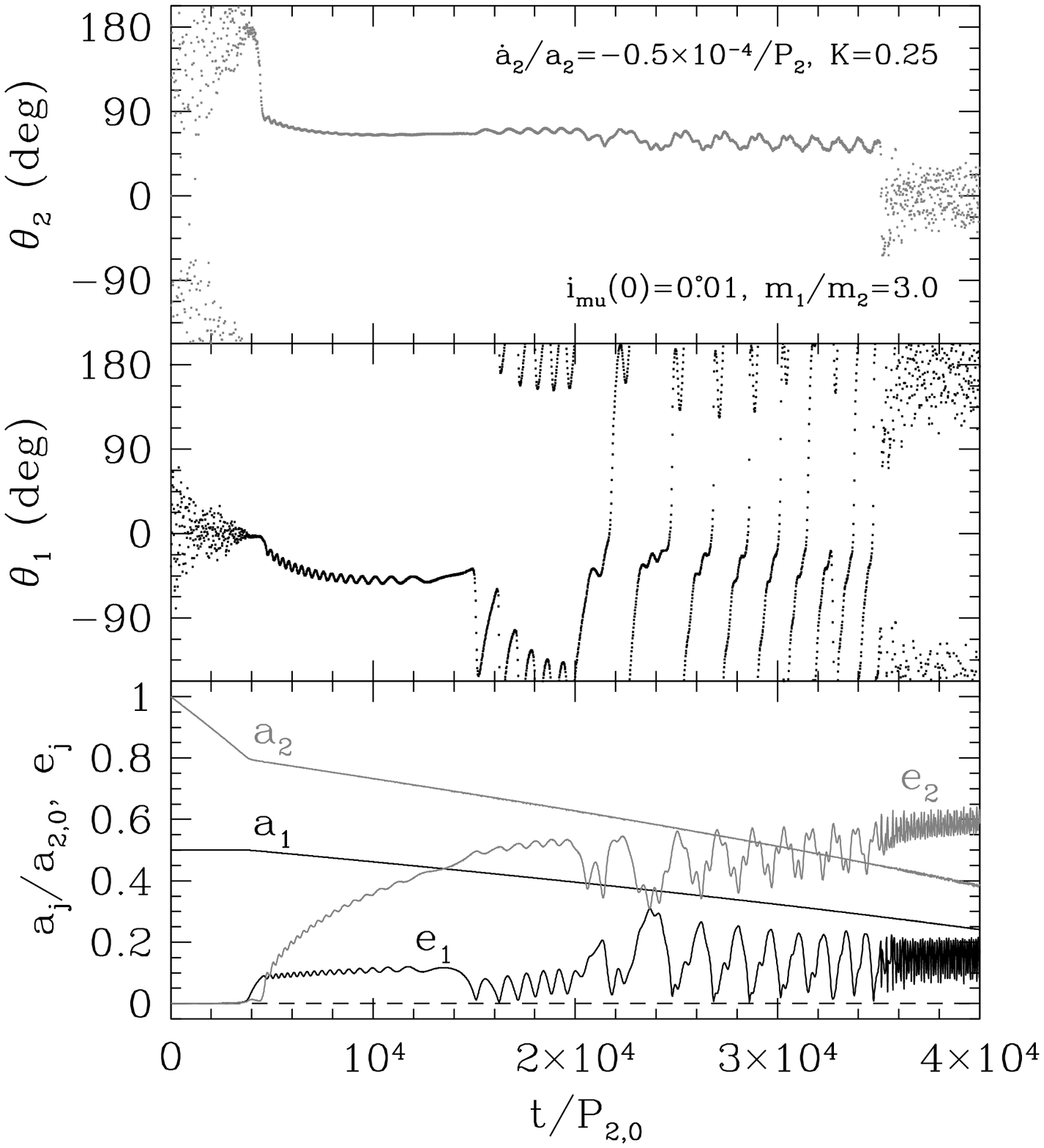}{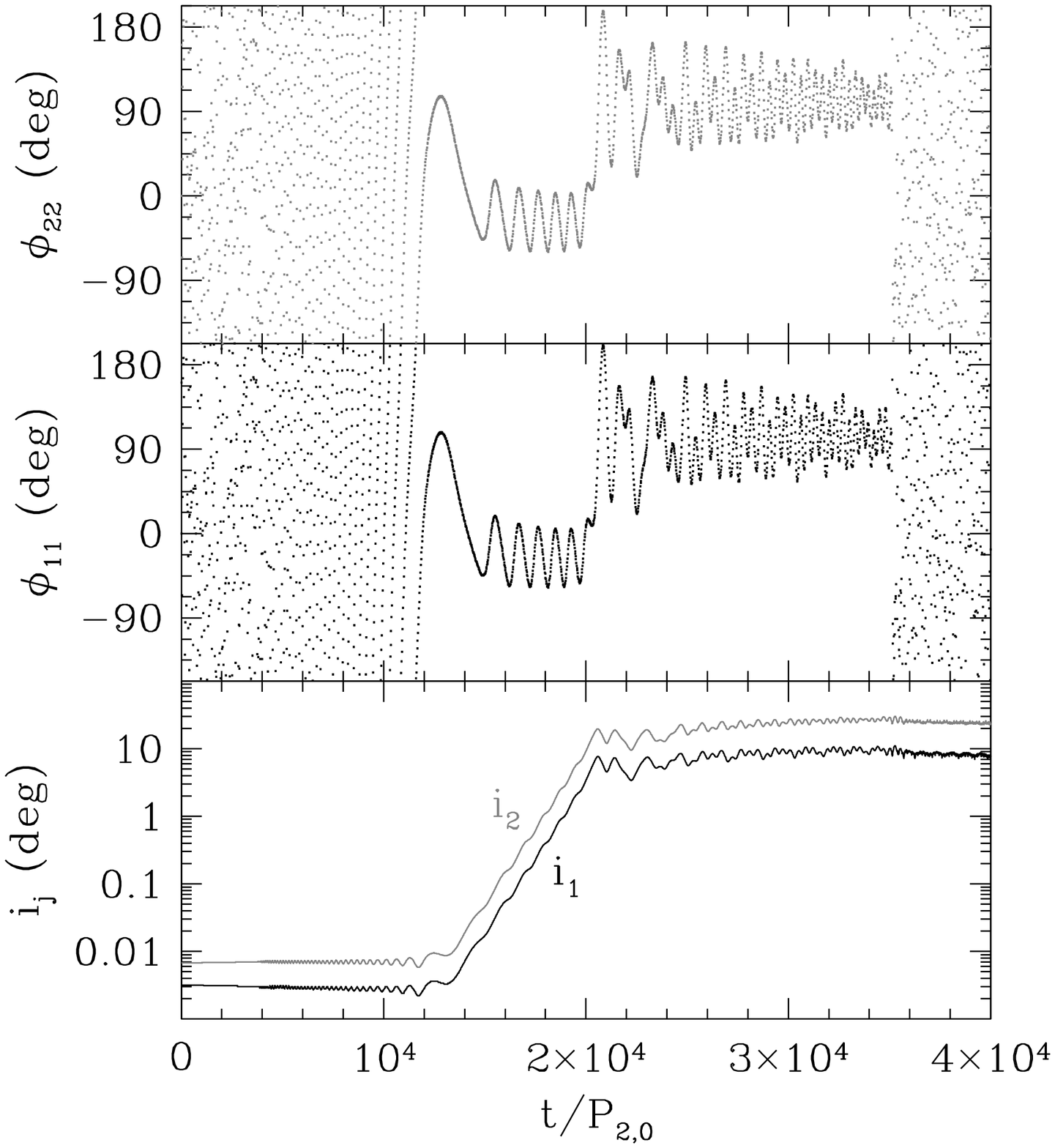}
\caption{\small
Same as Fig. \ref{fig:inclinedmed}, but with eccentricity damping
ratio $K = 0.25$.
In this case, the eccentricity damping results in clear libration of
$\phi_{jj}$ after the rapid inclination excitation phase.
\label{fig:edampmed}
}
\end{figure}

For faster migration rate, the effects of eccentricity damping on the
evolution of the system can be more complicated.
For example, the calculation shown in Figure \ref{fig:edampmed} is
similar to that in Figure \ref{fig:inclinedmed} ($m_1/m_2 = 3$ and
${\dot a}_2/a_2 = -0.5 \times 10^{-4} /P_2$) but with $K = 0.25$.
In this case, the eccentricity damping results in clear libration of
the inclination resonance variables $\phi_{jj}$ after the rapid
inclination excitation phase.
Nevertheless, the critical value of $K$ as a function of $m_1/m_2$
shown in Figure \ref{fig:edampsummary} also summarizes the results for
migration rate up to ${\dot a}_2/a_2 = -2 \times 10^{-4} /P_2$, if it
is interpreted as the critical value for inclination excitation, which
may or may not be followed by a phase with $\phi_{jj}$ clearly in
resonance.
For ${\dot a}_2/a_2 = -4 \times 10^{-4} /P_2$ (the maximum migration
rate studied), the critical value of $K$ is modified at large
$m_1/m_2$, with none of the calculations with $m_1/m_2 \ge 5$ showing
inclination excitation.

\section{CONCLUSIONS}
\label{conclusions}

We have investigated the effects of different migration rates on the
capture into and evolution in eccentricity and inclination resonances
at the 2:1 mean-motion commensurability by the convergent migration
of two planets.
We focused on systems with orbits that are initially slightly inclined
with respect to each other.
The system is first captured into the sequence I of 2:1 eccentricity
resonances found by \cite{lee04}, the same as in the case of exactly
coplanar orbits.
If the migration rate is fast and $m_1/m_2 \ga 2$, the subsequent
evolution is also identical to the coplanar case, with the
eccentricity resonances entering a new family (sequence IV), and there
is no inclination excitation or capture into inclination resonances.
The new family of 2:1 eccentricity resonances (with $e_2 \ga
0.4$--$0.5$, asymmetric librations, and orbits that intersect if they
are exactly coplanar) exists for $m_1/m_2 \ga 0.2$, but it is possible
to evolve into this family by fast migration only for $m_1/m_2 \ga 2$.
If the migration rate is slow, the system subsequently enters a phase
with the 4:2 inclination resonance variables $\phi_{jj}$ changing
slowly and the inclinations increasing rapidly, before it is captured
into 4:2 inclination resonances.
The inclination resonance configuration is symmetric, with $\phi_{11}
\approx \phi_{22} \approx 180^\circ$, if $m_1/m_2 \la 2.5$ and
asymmetric if $m_1/m_2 \ga 2.5$.
For intermediate migration rate (and fast migration rate if $m_1/m_2
\la 2$), there is typically a rapid inclination excitation phase, but
not a phase with $\phi_{jj}$ clearly in resonance.
We have also studied the effects of different eccentricity damping
rates during migration and found that the maximum value of $K =
|{\dot e}_2/e_2|/|{\dot a}_2/a_2|$ for inclination excitation (which
may or may not be followed by a phase with $\phi_{jj}$ clearly in
resonance if the migration rate is not slow) ranges from $\approx 0.7$
for $m_1/m_2 \ga 2.65$ to $\approx 2.8$ for $m_1/m_2 \sim 1$.
Since the evolution is sensitive to the rates of migration and
eccentricity damping within the ranges expected for type II
migration due to planet-disk interaction, the discovery of extrasolar
planetary systems with certain combinations of mass ratio and 2:1
resonance geometry would place a constraint on the strength of
eccentricity damping during migration, as well as on the rate of
migration itself.

There are several effects of disk-planet interaction that were
neglected in our analysis and may require further investigations.
We have focused on inward migration and eccentricity damping of the
outer planet, because previous hydrodynamic simulations (e.g.,
\citealt{kle00,kle04}) have shown that the disk inside the inner
planet's orbit, not just the disk material between the planets, should
be cleared rapidly.
However, \cite{cri08} have recently shown that a better numerical
treatment of the inner disk may result in a slower depletion of the
inner disk and that the eccentricity damping from the inner disk could
be important in explaining the observed eccentricities of resonant
pairs such as that in the GJ 876 system.
On the other hand, when the nearby disk mass is comparable to the
planet mass (i.e., in older, partially depleted disks), planets will
undergo type II migration at significantly less than the disk's
viscous advection speed \citep{sye95}, so that the inner disk
``outruns'' the planet and eventually leaves an inner hole, no matter
how small the inner boundary radius.
The simulations of \cite{tho08} suggest the majority of planets form
late enough in their parent disk's lifetime that such holes are
ubiquitous.

Planet-disk interaction can also affect the orbital inclination of a
planet.
The net effect of inclination damping by secular interactions and
excitation by interactions at mean-motion resonances depends on the
disk parameters, but any net damping should be on a timescale
comparable to or longer than the migration timescale (e.g.,
\citealt{lub01}).
This is likely too slow to affect the rapid inclination excitation
phase, but may result in equilibrium inclinations if the system is
subsequently captured into inclination resonances and the inclinations
are excited slowly by continued migration.

We have also neglected the secular apsidal and nodal precessions
induced by the disk, which could change the sequence of resonance
capture by changing and splitting the locations of the various
resonances at the same mean-motion commensurability.
\cite{kle05} have performed coplanar three-body integrations of the
GJ 876 resonant pair with additional apsidal precession and found that
the eccentricity resonances $\theta_1$ and $\theta_2$ are captured
into libration in a sequence that differs little in order or timing
from the case without additional apsidal precession.
This can be explained by the fact that the 2:1 eccentricity resonances
are first order, which means that the resonance-induced retrograde
apsidal precession is proportional to $1/e_j$ and much larger in
magnitude than the disk-induced prograde precession for small $e_j$.
On the other hand, disk-induced nodal precession could have a larger
effect, because the 4:2 inclination resonances are second order and
the resonance-induced nodal precession is roughly constant for small
$i_j$.
\cite{tho03} have performed some non-coplanar calculations with
additional apsidal and nodal precessions and did not find any
significant difference from the calculations without additional
precessions.
Although the adopted disk surface density is 5 times that of the
minimum mass solar nebula, they assumed an outer disk with an inner
edge that is likely too far (20 Hill radii) from the outer planet's
orbit, and the amount of precession induced by the disk is determined
primarily by the material closest to the planet.

\cite{ada08} and \cite{lec09} have recently examined the effects of
turbulence in circumstellar gas disks on mean-motion resonances in
extrasolar planetary systems.
They have found that stochastic perturbations due to turbulence could
prevent planets from staying in resonant configurations and that
planetary systems with mean-motion resonances should be rare.
This appears to be inconsistent with the observational evidence
discussed in \S \ref{intro}.
One possible explanation is that these studies assumed full
magnetorotational turbulence, whereas circumstellar disk models
usually exhibit an extensive dead zone around the midplane, where the
ionization fraction is low and the disk is magnetorotationally stable
due to ohmic dissipation (e.g., \citealt{gam96,san00,tur07,ilg08}).
For weak turbulence, the turbulence may generate larger libration
amplitudes than in smooth migration and allow, e.g., the jump from
sequence I to sequence IV to occur at a slower migration rate.

Finally, a better understanding of the capture into the inclination
resonances is needed.
Since the time spent in the phase with $\phi_{jj}$ changing slowly and
$i_j$ increasing rapidly is nearly independent of the migration rate
for slow migration (see \S \ref{noncoplanar}), in the limit of very
slow migration, there is an almost instantaneous jump in the
inclinations at the time of the inclination resonance capture, if we
measure time in units of the migration timescale.
This is not what one would expect if the capture into the inclination
resonances can be modeled by the appearance of additional equilibrium
points and separatrices in the Hamiltonian theory of a single
second-order resonance (see, e.g., \citealt{mur99}).
The Hamiltonian approach is based on the assumption that each
resonance is encountered individually, which is clearly not the case
in our problem.
In particular, the system is already in eccentricity resonances when
it enters the inclination resonances.
If the system is captured into just an inclination resonance, the
capture requires
${\dot \Omega}_j \approx -{\dot \lambda}_1 + 2 {\dot \lambda}_2$.
On the other hand, if the system is already in eccentricity
resonances so that
${\dot \varpi}_j \approx -{\dot \lambda}_1 + 2 {\dot \lambda}_2$,
then the capture into the inclination resonances requires
${\dot \omega}_j \approx 0$.

\acknowledgments
It is a pleasure to thank Stan Peale for informative discussions.
This research was supported in part by NASA grant NNG06GF42G (M.H.L.)
and a grant from NSERC Canada (E.W.T.).

\clearpage

\clearpage

\clearpage

\clearpage

\clearpage

\clearpage

\clearpage

\clearpage

\end{document}